# DIFNet: SAR RFI suppression based on domain invariant features

Fuping Fang[1,#], Wenhao Lv[2,#], Dahai Dai[1,*]

Affiliation 1: The School of Electronic Science, National University of Defense Technology.
Affiliation 2: School of Physics and Optoelectronic Engineering, Hangzhou Institute for Advanced Study, University of Chinese Academy of Sciences
# These authors contributed equally to this work.
* Corresponding author

*Abstract*—**Synthetic aperture radar is a high-resolution two-dimensional imaging radar, however, during the imaging process, SAR is susceptible to intentional and unintentional interference, with radio frequency interference (RFI) being the most common type, leading to a severe degradation in image quality. Although inpainting networks have achieved excellent results, their generalization is unclear, and whether they still work effectively in cross-sensor experiments needs further verification. Through time-frequency analysis of interference signals, we find that interference holds domain invariant features between different sensors. Therefore, this paper reconstructs the loss function and extracts the domain invariant features to improve the generalization. Ultimately, this paper proposes a SAR RFI suppression method based on domain invariant features, and embeds the RFI suppression into SAR imaging process. Compared to traditional notch filtering methods, the proposed approach not only removes interference but also effectively preserves strong scattering targets. Compared to PISNet, our method can extract domain invariant features and holds better generalization ability, and even in the cross-sensor experiment, our method can still achieve excellent results.**

## I. INTRODUCTION

SYNTHETIC aperture radar is an active microwave sensing system that adopts synthetic aperture and pulse compression techniques to acquire high-resolution images [1]. SAR systems with different bands are suitable for different applications. P-band SAR is commonly used for underground imaging and vegetation penetration. L, S, and C-band SARs are widely used for ocean monitoring and agricultural management. X and Ka-band SARs are often employed for high-resolution imaging [2,3]. During imaging, intentional and unintentional interferences often exist, with RFI being a widely common type, and the wide-range and high-intensity RFI significantly degrades SAR image quality [4,5].

To address these issues, numerous interference suppression algorithms have been proposed, broadly categorized into three types: non-parametric methods, parametric methods, and semi-parametric methods [6]. Regarding non-parametric methods, the letter [7] proposed an eigen-subspace-based filtering approach and this method holds very good compatibility with existing SAR imaging algorithms. The paper [8] proposed a generic subspace model for characterizing a variety of RFI types, and next designed a block subspace filter for removing RFI in SLC SAR. Parameterization methods often use iterative methods to solve interference parameters, and then filter out interference [9,10], which is often constrained by complex environment. The semi-parametric methods have gradually become mainstream due to its excellent performance, but they still face the drawback of high computational complexity. Common semi-parametric methods include sparse reconstruction [11], variants of robust PCA [12,13], and so on. Deep learning has been widely deployed in various fields due to its excellent performance [14,15], and naturally it is introduced into interference suppression [16,17]. The time-frequency domain radio frequency interference suppression method proposed in the literature [18] achieved better performance than robust PCA, and the networks proposed in [16,18] are collectively referred to as the image inpainting network.

Although image inpainting networks have achieved excellent results, their generalization is unclear, and whether they still work effectively in cross-sensor experiments needs further verification. What's more, in SAR interference suppression, there is a significant issue of incomplete data. Typically, we either obtain clean data or interfered data. Clean data and interfered data lack a corresponding relationship. To solve the above problems, this paper proposes a RFI suppression network based on domain invariant features, which offers the following contributions:

1. Through time-frequency analysis of interference signals, we find that interference holds domain invariant features between different sensors. Therefore, this paper reconstructs the loss function and extracts the domain invariant features to improve the generalization of the network. What's more, we also found that interference holds global characteristics on time-frequency map. Therefore, we adopt Transformer as the backbone network, and reduce the computational complexity by limiting the attention mechanism to local windows.

2. Compared to traditional notch filtering methods, our network avoids mistakenly classifying strong scattering targets,

and the proposed method achieves better interference suppression effect. Compared with image inpainting networks, this method holds stronger generalization ability in cross-sensor experiments. Even if the training data and testing data come from different sensors, the algorithm can still achieve excellent results. What's more, our method only requires the interfered data to perform interference suppression. Therefore, this approach can bypass the issue of incomplete data.

The organization of this paper is as follows: Section 2 introduces the signal model and the network. Section 3 presents the experimental results. And section 4 provides a summary of the entire paper.

## II. METHOD

This paper proposes a network based on domain invariant features (DIFNet), and embeds the RFI suppression into SAR imaging process. The overall process is illustrated in Fig. 1, and the algorithm is shown in Algorithm I.

| Algorithm I: DIFNet's pipeline |
| --- |
| 1. Detect RFI in SAR images; |
| 2. If there are SAR echoes: |
|     Locate interference echoes; |
|    else: |
|     Convert interfered images into echoes; |
| 3. Perform STFT pulse-by-pulse; |
| 4. Predict RFI by DIFNet; |
| 5. Subtract RFI; |
| 6. Perform ISTFT pulse-by-pulse; |
| 7. Convert SAR echoes into SAR images. |

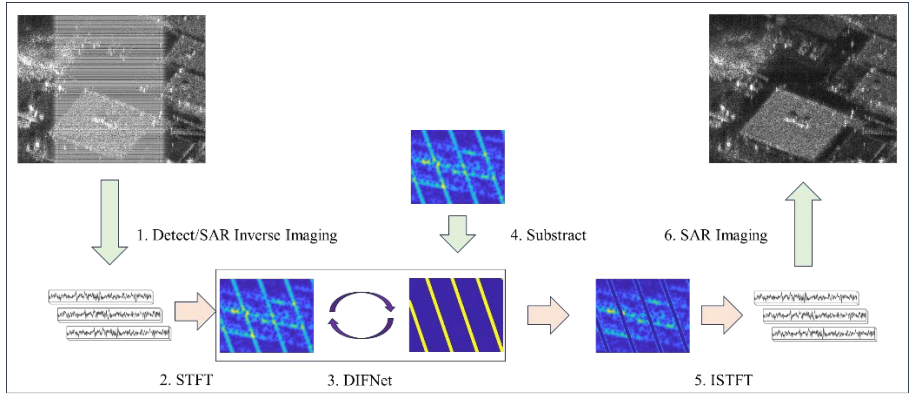

**Fig. 1.** Flow chart of RFI suppression network based on domain invariant features.

### A. RFI signal model

Common RFI can be categorized into narrowband interference, chirp broadband interference and sinusoidal broadband interference [6].

The narrowband interference can be expressed as:

$$s_{nbi}(\tau) = \sum_{n=1}^{N} A_n \exp\left(j2\pi f_c \tau + j2\pi f_n \tau\right) \quad (1)$$

Where, $f_c$ is the carrier frequency, $A_n$ is the amplitude, $f_n$ is the frequency offset, and $N$ is the number of interference signals.

The chirp modulation interference can be expressed as:

$$s_{wbi_{cm}}(\tau) = \sum_{n=1}^{N} B_n \exp\left(j2\pi f_c \tau + j\pi k_n \tau^2\right) \quad (2)$$

Where, $B_n$ is the amplitude, and $k_n$ is the chirp rate.

The sinusoidal modulation interference can be represented as:

$$s_{wbi_{sm}}(t) = \sum_{n=1}^{N} C_n \exp\left(j2\pi f_c t + \beta_n \sin\left(2\pi f_n t\right)\right) \quad (3)$$

Where, $C_n$ is the amplitude, $\beta_n$ is the modulation coefficient and $f_n$ is the modulation frequency. According to the Bessel formula, sinusoidally modulated broadband interference can be expressed as:

$$s_{wbi_{sm}}(t) = \sum_{n=1}^{N} \left( C_n \exp\left(j2\pi f_c t + \beta_n \sin\left(2\pi f_n t\right)\right) \right)$$
$$= \sum_{n=1}^{N} \left( C_n \sum_{m=-\infty}^{m=\infty} \binom{J_{2m}(\beta_n)\sin\left(2\pi f_c t + 2\pi m f_n t\right)}{+ J_{2m}(\beta_n)\cos\left(2\pi f_c t + 2\pi m f_n t\right)} \right) \quad (4)$$

Where, $J_{2m}(\beta_n) = \sum_{k=0}^{\infty} \dfrac{(-1)^k}{k!(m+k)!}\left(\dfrac{\beta_n}{2}\right)^{m+2k}$. According to (4), sinusoidally modulated broadband interference is composed of a series of narrowband interferences, approximating a specific form of chirp modulated broadband interference.

Formulas (1), (2), (4) can be uniformly expressed as:

$$s_{RFI}(t) = \sum_{n=1}^{N} D_n \exp\left(j2\pi f_c t + j\pi k_{RFI} t^2\right) \quad (5)$$

Where, $D_n$ is the amplitude, $k_{RFI}$ is the tuning rate. When $k_{RFI}$ is small, RFI is a narrowband interference, and when $k_{RFI}$ is large, RFI is a broadband interference.

By conducting time-frequency analysis on equation (5), it can be found that:

$$f = k_{RFI} \cdot t \quad (6)$$

From formula (6), it can be seen that the signal holds global characteristics on the time-frequency map, which inspires us to propose a Transformer network. Moreover, from the interference model in Equation (5), it can be seen that the interference signal does not vary with different radar sensors. Therefore, this inspires us to extract domain-invariant features of the interference to improve generalization.

### B. Network

Fig. 2 illustrates the overall architecture of Transformer's variant, which is a U-shaped network. The input to the network is the interfered image $I \in R^{1 \times H \times W}$, and the output is the label $O \in R^{1 \times H \times W}$. The Transformer network consists of an encoder, feature extraction module, and decoder. In the encoder, bottom-level features are extracted by a $3 \times 3$ convolutional layer and LeakyReLU activation, and the extracted feature can be expressed as $X_0 \in R^{C \times H \times W}$. Following the U-shaped architecture, the feature is sequentially processed by $L$ encoders, each containing multiple LeWin Transformers and a downsampling layer. The LeWin Transformer consists of a non-overlapping window self-attention block and a LeFF network. The downsampling layer reduces the length and width of the input by half and doubles the number of channels. For a given input $X_0 \in R^{C \times H \times W}$, the output feature map of the $l$-th stage can be represented as $X_l \in R^{2^l C \times \frac{H}{2^l} \times \frac{W}{2^l}}$. The bottom-level feature extraction module in Fig. 2(a) consists of multiple LeWin Transformer blocks, which leverage the hierarchical structure to capture longer-range information and even global information. The decoder also consists of $L$ stages, each containing an upsampling layer and multiple LeWin Transformer blocks. Where the upsampling layer is a deconvolutional layer with a stride of 2, increasing the feature resolution while reducing the number of channels. The upsampled features and the features from the same-level encoder are then fed into the LeWin Transformer block. After $L$ decoder stages, the features are mapped to the original image size by an output projection layer.

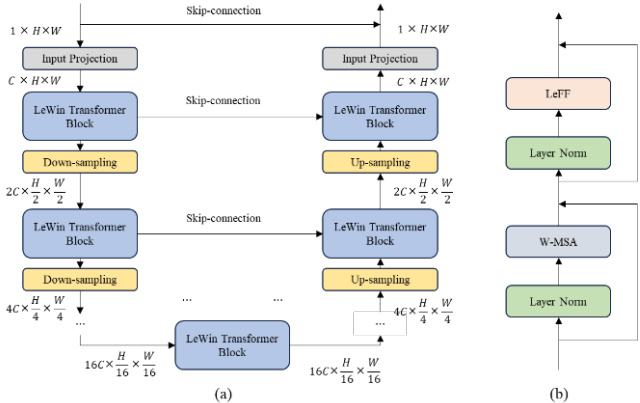

**Fig. 2.** DIFNet structure. (a) The overall structure of the network. (b) LeWin Transformer Block.

### Module-1: LeWin Transformer Block

The LeWin Transformer Block consists of W-MSA (Window-based Multi-Scale Self-Attention) and LeFF (Local Enhanced Feed-Forward) modules, which offer the following advantages. First, compared to traditional Transformers, it significantly reduces computational complexity. Standard Transformers compute global self-attention across the entire image, resulting in high computational complexity. While our method utilizes a non-overlapping local window attention module, which greatly reduces the computational burden. Second, this module can capture both global and local contextual information. W-MSA captures long-range information, while LeFF captures local information.

The LeWin Transformer block can be represented as:

$$X_l^{'} = \text{W-MSA}\big(\text{LN}\big(X_{l-1}\big)\big) + X_{l-1} \tag{5}$$

$$X_l = \text{LeFF}\big(\text{LN}\big(X_l^{'}\big)\big) + X_l^{'} \tag{6}$$

### Module-2: W-MSA

Unlike the global self-attention mechanism in standard Transformers, the non-overlapping local window self-attention mechanism efficiently reduces computational complexity. We first split $X \in R^{C \times H \times W}$ into $M \times M$ non overlapping blocks, and the input data to W-MSA is $X^i \in R^{C \times M \times M}$. Next, we perform self-attention on each window. The computation process for the $k$-th head can be described as follows:

$$X = \big\{X^1, X^2, \cdots, X^N\big\}, N = HW / M^2 \tag{7}$$

$$Y_k^i = Attention\big(X^i W_k^Q, X^i W_k^K, X^i W_k^V\big), i = 1, \cdots, N \tag{8}$$

$$\hat{Y}_k = \big\{Y_k^1, Y_k^2, \cdots, Y_k^1\big\} \tag{9}$$

The results of all heads are concatenated and linearly projected to obtain final result. In addition, inspired by the Swin Transformer, we introduced relative position encoding $B$ in the attention module. Therefore, the attention calculation process can be expressed as follows:

$$Attention(Q, K, V) = \text{Soft} \max\left(\frac{QK^T}{\sqrt{d_k}} + B\right)V \tag{10}$$

Compared to global self-attention mechanism, the computational complexity of W-MSA has been reduced from $O\big(H^2 W^2 C\big)$ to $O\big(M^2 HWC\big)$, where $M$ is the window size and is generally much smaller than $H$ and $W$. Therefore, the final computational complexity is significantly reduced.

### Module-3: LeFF

The standard Transformer's feed-forward network has limited capability in capturing local contextual information. Considering the importance of neighboring pixels for image tasks, we introduce a deep convolutional module in the feed-forward network to capture local contextual information.

### Module-4: Loss function

In inpainting network, the loss can be defined as:

$$l = \sqrt{(HX - Y)^2 + \varepsilon} \tag{11}$$

$H$ is the network's matrix, $X$ is the input, and $Y$ is the label. In order to extract domain invariant interference features, $Y$ was optimized as follows:

$$Y_{pixel} = \begin{cases} 1, pixel \in RFI \\ 0, pixel \notin RFI \end{cases} \tag{12}$$

Pixel represent pixel value of the image, and in equation (12), it can ensure that the measurement distance of RFI does not change in different sensors. Therefore, It can induce the network to learn domain invariant features, thereby ignoring targets.

So, the loss can be expressed as:

$$l = \sqrt{\left(HX_{pixel} - Y_{pixel}\right)^2 + \varepsilon} \qquad (13)$$

## III. EXPERIMENTS

To validate the effectiveness of the proposed method, experiments are conducted on both MiniSAR dataset and Sentinel-1 dataset. The resolution of MiniSAR is 0.1m, while the resolution of Sentinel-1 is 5×20m. All training data comes from MiniSAR, and the testing data comes from MiniSAR and Sentinel-1. The cross-sensor and cross-resolution experiments can be used to verify the generalization.

### A. Evaluation Metrics

To reasonably evaluate the test results, this paper adopts pixel accuracy (PA), intersection over union (IoU), PSNR [18] and ME [12] as evaluation metric. PA and IoU are used to evaluate the DIFNet, and PSNR and ME is used to evaluate image quality.

ME is defined as follows:

$$ME = \text{Ent}\left(\hat{X}\right)\text{Mean}\left(\hat{X}\right) \qquad (14)$$

$\text{Ent}\left(\hat{X}\right)$ is the entropy, $\text{Mean}\left(\hat{X}\right)$ is the mean value. A smaller entropy indicates that the pixel values of the image are concentrated within a smaller range. A smaller mean value indicates a lower amplitude, suggesting that most of the interference has been filtered out. Therefore, a smaller ME indicates a better filtering result.

### B. MiniSAR

The filtering results on the time-frequency map are shown in Fig. 3. Comparing Fig. 3(b) and Fig. 3(d), it can be observed that for the traditional notch filtering method, some strong scattering targets are mistakenly filtered out as interference, but for DIFNet, these strong scattering targets are well preserved. Where, the image restoration network PISNet achieved the best result. Fig. 4 shows the imaging results, and it can be seen that the proposed algorithm preserves more details and produces a cleaner filtering result comparing with traditional notch filter. The evaluation results are presented in TABLE I, it can be seen that the proposed method achieves a 1.6% improvement in PA, a 5.99% improvement in IoU, a 2.02dB improvement in PSNR, and a 0.05 decrease in ME compared to the traditional algorithm. PISNet still achieved the best result.

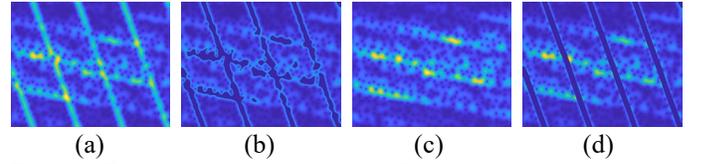

**Fig. 3.** Time frequency map; (a) Interfered time-frequency map; (b) Time frequency map by notch filtering; (c) Time frequency map by PISNet; (d) The time-frequency map by DIFNet.

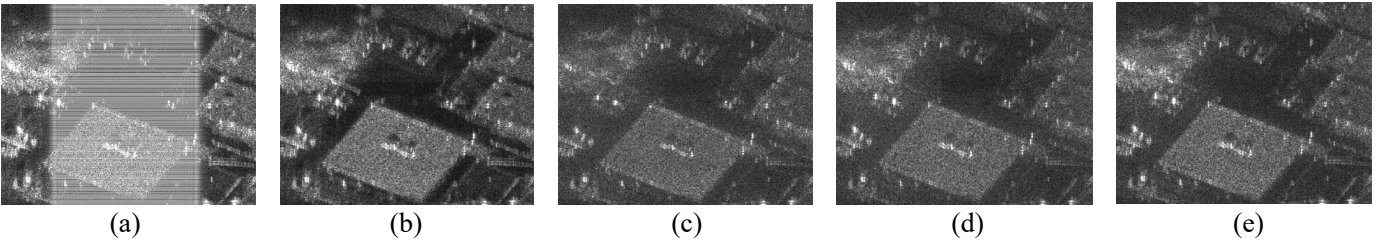

**Fig. 4.** Suppression results in MiniSAR; (a) Interfered image; (b) Label; (c) Result by notch filtered; (d) Result by PISNet; (e) Result by DIFNet.

TABLE I
SIMUTATION RESULT

| Parameters | Interfered image | Notch filtering | PISNet | DIFNet |
|---|---|---|---|---|
| PA | 83.09% | 94.95% | / | 96.55% |
| IoU | / | 74.41% | / | 80.40% |
| PSNR | / | 21.49 | **25.05** | 23.51 |
| ME | 0.24 | 0.09 | **0.04** | **0.04** |

### C. Sentinel-1

The interfered dataset is obtained from Sentinel-1, captured in the Korean region on 2/16, 2019. The image area is cropped to a size of 512×512. The time-frequency map is shown in Fig. 5, and the filtering results and performance indicators are shown in Fig. 6. In the Sentinel-1 testing datasets, the training data still comes from MiniSAR. And in this cross-sensor experiment, PISNet does not work, so we do not present its results, while our method can still suppress the interference, demonstrating its good generalization. Comparing Fig. 5(b) and Fig. 5(c), it can be observed that the proposed method achieves a better filtering result. Similarly, comparing Fig. 6(b) and Fig. 6(c), it can be seen that low-intensity residual interference is difficult to filter out for traditional notch filter, while our method can effectively filter out the residual interference. Comparing with traditional notch filter, our method has reduced ME by 0.15.

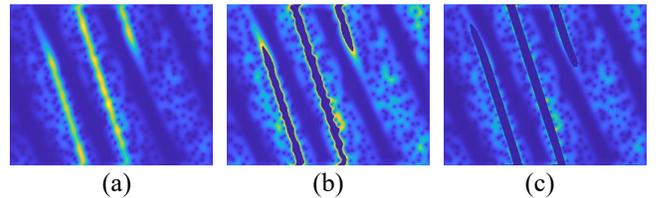

**Fig. 5.** Time-frequency map; (a) Interfered time-frequency map; (b) Notch filtering result; (c) DIFNet filtering result.

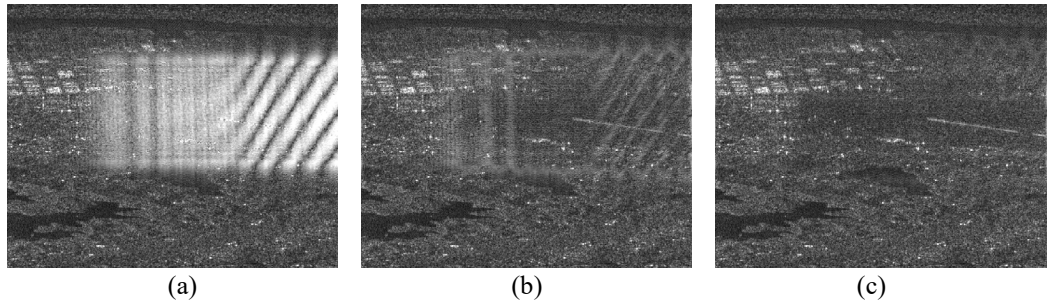

| (a) | (b) | (c) |

**Fig. 6.** Suppression results in Sentinel-1; (a) Interfered image, *ME*=3.40; (b) Notch filtering image, *ME*=2.34; (c) DIFNet filtering image, ***ME*=2.19**.

From the above experimental results, it can be seen that in the cross-sensor experiment, the image inpainting network does not even work, but our method can still achieve excellent results. The above results demonstrate that our method holds good generalization.

## IV. CONCLUSION

SAR is widely deployed as a high-resolution imaging radar, but it is susceptible to intentional and unintentional RFIs. For image inpainting networks, although they have achieved excellent results, their generalization is unclear. To address this problem, through time-frequency analysis of interference signals, we find that interference holds domain invariant features between different sensors, and propose a SAR RFI suppression network based on domain invariant features. Compared to traditional notch filtering methods, the proposed method achieves better interference suppression effect. Furthermore, in the cross-sensor experiment, the training data and the testing dataset comes from different radars with different resolutions, and the image inpainting networks does not work, but our method can still achieve excellent results. The above demonstrates that our method holds good performance and generalization. Moreover, this method can inspire self-supervised learning, as the segmented time-frequency map forms a masking task, which can be repaired by self-supervised networks.

## ACKNOWLEDGMENT

Thank you to the School of Electronic Science, National University of Defense Technology for your support of this project.